# The Quantum Links of the Anyons and their Entanglement

Viorel Laurentiu Cartas

In this paper, after a brief presentation of the physical 2+1 dimensional place where the anyons evolve, there is established the link's creation probability for the anyons. The departure point is the celebrated Laughlin wave function.

Then, two cases of quantum links are emphasized, considering two and three anyons. In the first case a two qubits register is attached. The unitary transformation (e.g. the $U_{CNOT}$ unitary transformation) can map not-entangled quantum states to the quantum and topological entangled states, as there are highlighted in this paper for the two anyons link case.

The six qubit register which is attached to the three anyons link is only shortly presented since it represents a work in progress. The topologic and quantum states are represented by ket vectors and correspondingly by diagrams.

*Introduction*

The evolution of the particle in the 2+1 space (two spatial dimensions and one temporal dimension) is very different from the particle that evolved in the 3+1 space. The latter obeys two extreme statistics: the Bose-Einstein statistics (being boson and since the bosons may simultaneously occupy the same quantum states) and the Fermi-Dirac statistics (being fermion and since two fermions are not allowed to occupy the same quantum state). The former is special in many aspects. It obeys a particular statistic named fractional statistic. Such particles are called anyons (the reason is obvious) and they are allowed to occupy the same quantum state in a certain number (number of occupation).

In the 3+1 space there are not supplementary constrains in order to interchange two particles. It is not the case with the 2+1 space, when the two particles have two different ways to interchange: either on the left side or on the right side.

In the last thirty years this subject has become a major issue, succeeding to generate modern ideas in a large range of domains such as: the braiding algebra, the knot theory, the quantum computing (with its sub domain named topological quantum computation that is more and more explored nowadays [1], [2]),etc.

In the same time the hunting of other new quasi particles that are supposed to evolve in 2+1 dimensions space have determined the developing of important experiments concerning the two dimensional electron systems:

- MOSFET (Metal Oxide Semiconductor Field Effect Transistor) using the effect of the inversion layer
- Super lattice which is basically a two dimensional electron system that appears in the heterostructures of two semiconductors
- Liquid Helium Surface where two contrary actions coexist, constraining the particles to remain in two spatial dimensions (the potential barrier and the mirror potential).

In certain situations (high magnetic field and very low temperature) in the first and second examples the Quantum Hall effect does act. The study of the QHE has begun with the seminal papers of K. von Klitzing [3] in 1980 for the Integer Quantum Hall effect and D.C.Tsui [5] in 1982 for Fractional Quantum Hall effect. The anyons are the quasi particles which represent the objects of the effects. They have fractional electron charges.

R.B.Laughlin[3] is the person who has studied FQHE theoretically, earning the Nobel Prize for his work. He has introduced the Landau level filling factor $v$ = p/m with *p,m* integers. More, Laughlin has introduced (on the extreme quantum limit in which the Landau level hierarchy is large enough, allowing that all particles are found in the lowest Landau level) a single particle wave function:

$$\psi_m(z) = C \cdot z^m \cdot e^{-\frac{1}{4}|z|^2} \qquad (1).$$

For the many particle case, Laughlin has introduced a wave function corresponding only for the lowest Landau level:

$$\psi_m(z) = \prod_{i<j}(z_i - z_j)^m e^{-\frac{1}{4}\sum_k |z_k|^2} \qquad (2),$$

where *m* is an odd integer. The Laughlin idea is based on a bound liquid droplet model.

*The anyons' interchange probability*

This paper is searching for a satisfactory response to an important question about the anyons. Taking into account the Laughlin wave function (1) as the proper function that can describe the anyon acting in a physical space miming a 2+1 dimensions space, it is mandatory to consider the collapse of the wave function as well. Consequently, at a certain moment in time the corresponding projections of the waves on the two dimensional space will intersect.

So, there is a nonzero probability to find anyons in the same space. The good question is: which is the probability to find the anyons interchanged, when we make a measurement?

From the macroscopic point of view, there are two different possibilities to interchange the anyons: either on the right or on the left. The corresponding probabilities are equal. From the quantum point of view, there are these two possibilities to interchange (we noted them with |R> and |L>) and many others obtained as superposition (a|R> +b|L> if $|a|^2+|b|^2=1$).

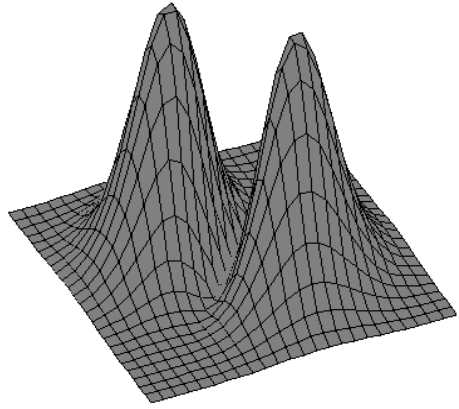

Fig.1
The Laughlin single particle wave functions

In Fig.1 we can see the shape of the wave function described by the expression (1) were m=3. In Fig.2 we have considered two anyons described by there wave packets. The radius of the circular projection of the anyon's wave packet on the two dimensional space is the uncertainty $\Delta x(t)$ of the anyon's position. The width of the packets grows in time and the pulses collapse. Different initial shape of the anyon's pulse determines different ways to collapse .For instance in Fig 2 the first anyon pulse evolves from a large contour ( comparing to the wave width) to a collapsed but almost unchanged one. The collapsing velocity is small. On the contrary, the second anyon being initially represented by a high amplitude and thin pulse has a high collapsing velocity.

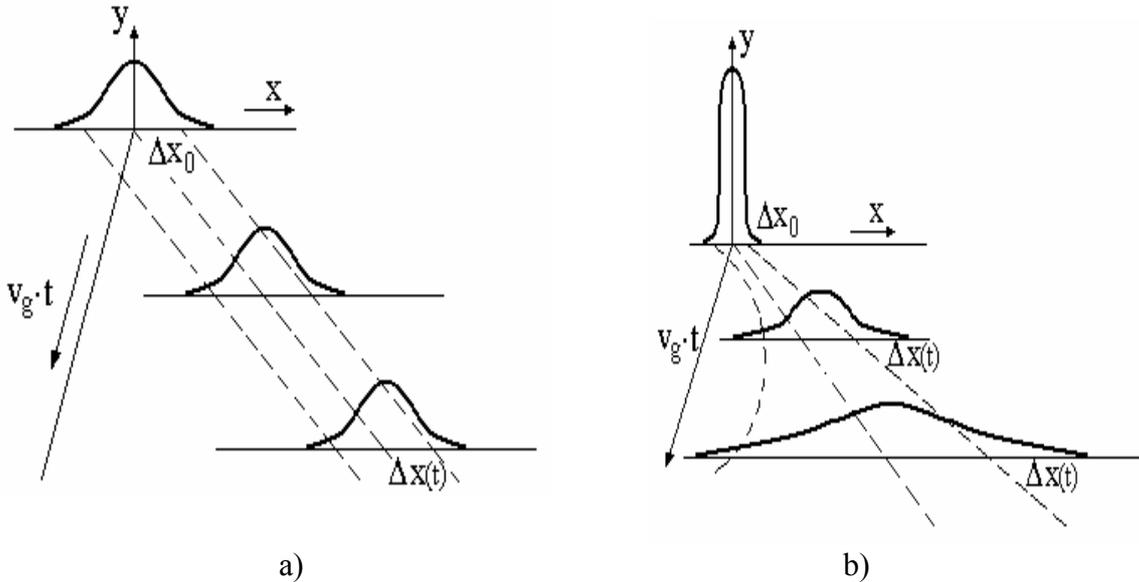

a)  b)

Fig.2
Two anyons described by there wave packets; the collapsing velocity is small a) or high b) due to the wave packet's shape

The position uncertainty as a time function is:

$$\Delta x(t) \cong \sqrt{(\Delta x_0)^2 + \frac{k \cdot t^2}{\Delta x_0^2}} \qquad (3),$$

where $\Delta x_0$ is the initial position uncertainty of the anyon; $k$ is constant. For a large $t$ we assume that $\Delta x(t)$ is linear.

Considering all these, we can see that if we'll create the necessary experimental conditions in order to determine the anyons to evolve dynamical, sooner or later, special areas will inevitable appear. If we'll make a measurement in those areas we'll be able to find multiple anyons (every anyon with its specific probability to exist there).

Specifically, considering only two anyons well defined in the two dimensional space, they will share (after a certain time) an area, where they may be found together. They may be found in the same relative position or in the inverse position (one relative to the other). In the second case, there are not only two different ways to change the relative position (on the left or on the right), but many others (in fact, any acceptable superposition).

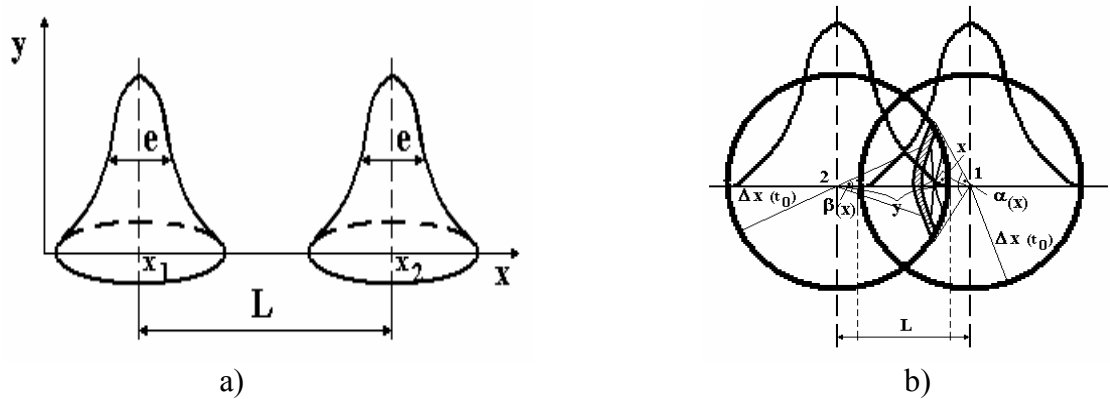

Fig.3

a) The anyons wave packet's shape and b) their projections on 2 dimensions space

If the anyons are separated by L distance, the probability to find one of those anyons in the shared area is

$$P_1 = P_2 = \int_{L-\Delta x_0}^{\Delta x_0} x \cdot \alpha(x) \cdot |\psi(x)|^2 dx \qquad (3)$$

where $\psi(x)$ is the Laughlin wave function (1).
We use the next relation in order to norm:

$$\int_0^{\Delta x_0} x \cdot \alpha(x) \cdot |\psi(x)|^2 dx = \frac{1}{2} \qquad (4).$$

The probability to find the anyon "2" in the shared area on the right side of the shadowed strip (corresponding to the area where the anyon "1" may be found) is:

$$P_{2_R} = \int_{L-x}^{\Delta x_0} y \cdot \beta(x) \cdot |\psi(y)|^2 dy \qquad (5).$$

The similar probability for the anyon "2" for the left side, is:

$$P_{2_L} = \int_{L-\Delta x_0}^{L-x} y \cdot \beta(x) \cdot |\psi(y)|^2 dy \qquad (6);$$

$\alpha(x)$ and $\beta(x)$ represent the correspondent angles of the arcs where the anyons could be found, when a measure took place (Fig.4).

The anyon "1" may be exactly on the shadowed strip when the anyon "2" is in the right side of the strip (Fig.4) with the probability:

$$P_{2_R} \cdot dP_1 = \left[ \int_{L-x}^{\Delta x_0} y \cdot \beta(x) \cdot |\psi(y)|^2 dy \right] \cdot x \cdot \alpha(x) \cdot |\psi(x)|^2 dx \qquad (7).$$

Now, in the absence of any asymmetric field, the $\Pi$ probability for the anyon "1" to interchange the anyon "2" on the right side is the same with the probability for the anyon "1" to interchange the anyon "2" on the left side:

$$\Pi = \frac{1}{2} \int_{L-\Delta x_0}^{\Delta x_0} \left[ \int_{L-x}^{\Delta x_0} y \cdot \beta(x) \cdot |\psi(y)|^2 dy \right] \cdot x \cdot \alpha(x) \cdot |\psi(x)|^2 dx \qquad (8).$$

With $\alpha(x)$ and $\beta(x)$, being:

$$\begin{cases} \alpha(x) = 2\arccos \dfrac{(x + \Delta x_0 + L)^2 - 4(\Delta x_0^2 - x^2)}{4x(x + \Delta x_0 + L)} \\ \\ \beta(x) = 2\arccos \dfrac{(x + \Delta x_0 + L)^2 + 4(\Delta x_0^2 - x^2)}{4\Delta x_0(x + \Delta x_0 + L)} \end{cases} \qquad (9),$$

the $\Pi_0$ probability corresponding to the initial moment, is:

$$\Pi_0 = 2C \int_{L-\Delta x_0}^{\Delta x_0} \int_{L-x}^{\Delta x_0} x^7 y^7 e^{-\frac{1}{2}(x^2+y^2)} \cdot \arccos \frac{(x+\Delta x_0+L)^2 - 4(\Delta x_0^2 - x^2)}{4x(x+\Delta x_0+L)} \cdot \arccos \frac{(x+\Delta x_0+L)^2 + 4(\Delta x_0^2 - x^2)}{4\Delta x_0(x+\Delta x_0+L)} dy \cdot dx$$

(10),

where $C$ is a constant which may be found from (4). For instance, with proper unities, if $\Delta x_0 = 1$ and L=1.75 then $\Pi = 2C \cdot 0.383$.

*The quantum-link states and their diagrams*

From the quantum point of view there are two possibilities to interchange (on the

right side 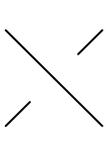 or on the left side 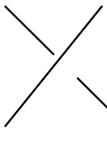 and we note the correspondent states with $|R\rangle$ and $|L\rangle$) and many others, obtained as superposition in Hilbert space ($|\psi\rangle = a|R\rangle + b|L\rangle$, with $|a|^2 + |b|^2 = 1$; $a,b \in C$ ).We represent the $|\psi\rangle$ state as 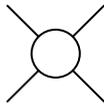 and we'll call it as *the quantum-link state*. This is similar to a qubit; actually it is a qubit. The consecutive interchange is represented by concatenating the crossovers. A quantum-link state may have different registers of qubits. So, a quantum-2-link has a register of two qubits but a quantum-3-link has a register of four qubits(if there exist two of three annuli that do not make a link between them) or has a register of six qubits (if every annulus make links with all the other annulus). The analyze is more complex for greater order.

Both states $|R\rangle$ and $|L\rangle$ are orthogonal. So, $\langle R|R\rangle, \langle L|L\rangle \neq 0$ this means that

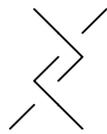 and 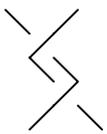 do not equal zero and $\langle R|L\rangle = 0$, which

means that 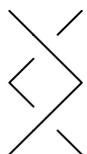 equals 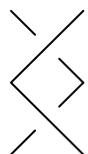 and equals 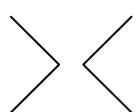 ,and all of them equal zero.

For the quantum-link state, the density operator is $\rho = |\psi\rangle\langle\psi|$ and it has the matrix density:

$$[\rho_{kj}] = [\langle k|\rho|j\rangle] = \begin{bmatrix} |a|^2 & ab^* \\ a^*b & |b|^2 \end{bmatrix} \qquad (11).$$

Considering an operator $O$ that corresponds to an observable, the expected value is:

$$\langle\psi|O|\psi\rangle = Tr(\rho O) = \sum_k \sum_j \rho_{kj} A_{jk} \qquad (12).$$

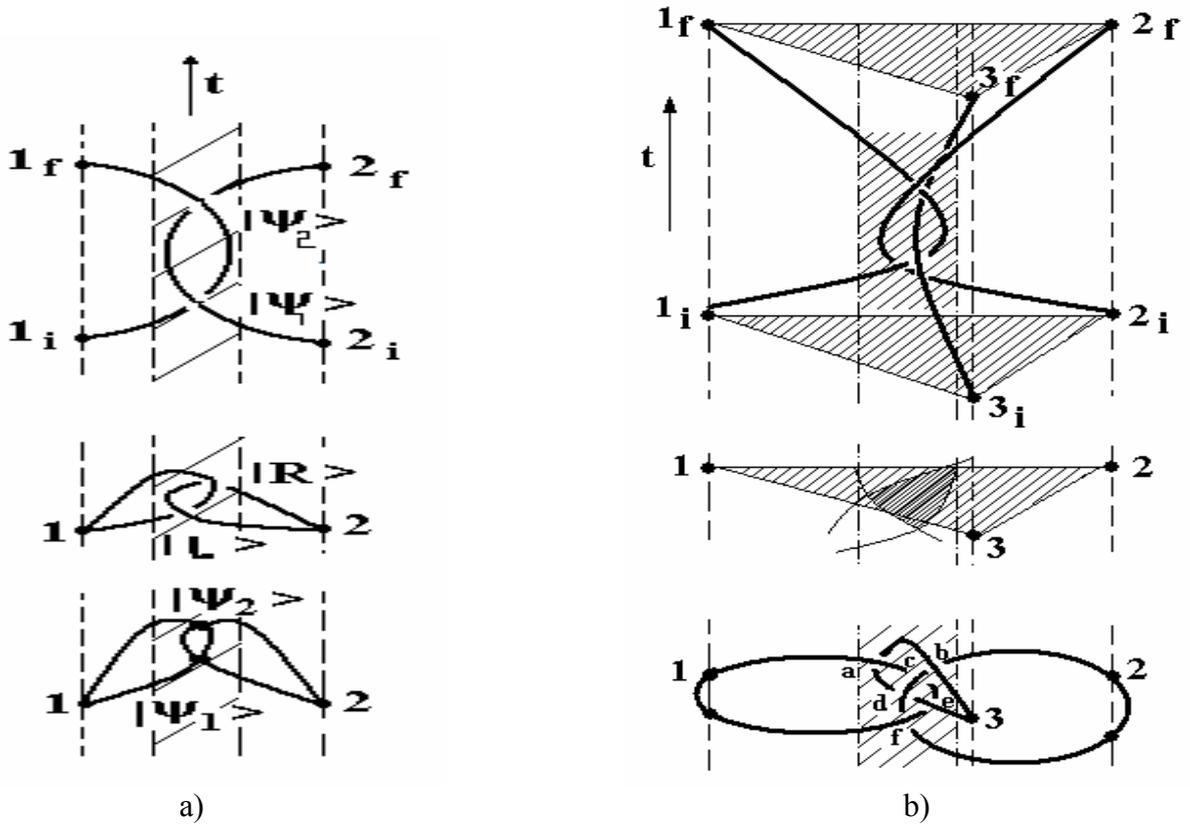

a) b)

Fig.4.

a) The quantum-2-link with two qubits register (the quantum braid and the quantum link of two anyons), b) The quantum-3-link with six qubits register (the quantum braid and the quantum link of three anyons)

For the quantum-2-link (which is the simpler one) (Fig.4, a)), we assume that at $t_i$ there are two anyons and they interchange their places twice, till the final moment $t_f$. So, do appear two quantum-link states $|\psi_1\rangle$ and $|\psi_2\rangle$:

$$|\psi_1\rangle = a_1|R\rangle + b_1|L\rangle \ , \ |\psi_2\rangle = a_2|R\rangle + b_2|L\rangle \qquad a_1, b_1, a_2, b_2 \in C \ ;$$

thus, there it is a two qubits register.

1). If $a_1 = 1, b_1 = 0$ then, $[\rho_{kj}] = \begin{bmatrix} 1 & 0 \\ 0 & 0 \end{bmatrix}$. If $a_2 = 1, b_2 = 0$ then, $[\rho_{kj}] = \begin{bmatrix} 1 & 0 \\ 0 & 0 \end{bmatrix}$. So, we have the same $[\rho_{kj}]$.

2). If $a_1 = 0, b_1 = 1$ then, $[\rho_{kj}] = \begin{bmatrix} 0 & 0 \\ 0 & 1 \end{bmatrix}$. If $a_2 = 0, b_2 = 1$ then, $[\rho_{kj}] = \begin{bmatrix} 0 & 0 \\ 0 & 1 \end{bmatrix}$. Again, we have the same $[\rho_{kj}]$.

3). If $a_1 = 1, b_1 = 0$ then $[\rho_{kj}] = \begin{bmatrix} 1 & 0 \\ 0 & 0 \end{bmatrix}$. If $a_2 = 0, b_2 = 1$ then $[\rho_{kj}] = \begin{bmatrix} 0 & 0 \\ 0 & 1 \end{bmatrix}$. So, we don't have the same $[\rho_{kj}]$.

4). If $a_1 = 0, b_1 = 1$ then $[\rho_{kj}] = \begin{bmatrix} 0 & 0 \\ 0 & 1 \end{bmatrix}$. If $a_2 = 1, b_2 = 0$ then $[\rho_{kj}] = \begin{bmatrix} 1 & 0 \\ 0 & 0 \end{bmatrix}$. Again, we don't have the same $[\rho_{kj}]$.

1) and 2) have the same topology; the annuli are linked. 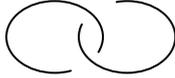 equals 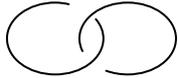

In 3) and 4) the annuli are not linked. It is similar with the situation when the two anyons do not change their places. 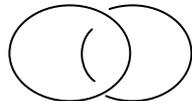 equals 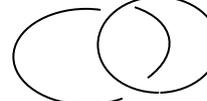 equals 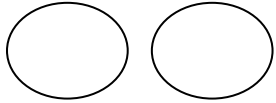.

In Fig.4, b) we have considered three anyons and this goes to a register of six qubits. In a general way for a register of *n* qubits we consider a two dimensional Hilbert space *H* and an arbitrary orthonormal base $\{|0\rangle, |1\rangle\}$. We are taking into account *n* different Hilbert spaces, $H_{n-1}$, $H_{n-2}$,…, $H_0$ isomorphic with *H* having the bases:

$$\{\{|0_{n-1}\rangle, |1_{n-1}\rangle\}, \{|0_{n-2}\rangle, |1_{n-2}\rangle\}, ..., \{|0_0\rangle, |1_0\rangle\}\}.$$

Considering that we have prepared those n qubits in the states:

$$2^{-\frac{1}{2}}(|0_{n-1}\rangle + |1_{n-1}\rangle), 2^{-\frac{1}{2}}(|0_{n-2}\rangle + |1_{n-2}\rangle), ..., 2^{-\frac{1}{2}}(|0_0\rangle + |1_0\rangle).$$

The $|\psi\rangle$ state is a product tensor:

$$|\psi\rangle = 2^{-\frac{1}{2}}(|0_{n-1}\rangle + |1_{n-1}\rangle) \otimes 2^{-\frac{1}{2}}(|0_{n-2}\rangle + |1_{n-2}\rangle) \otimes ... \otimes 2^{-\frac{1}{2}}(|0_0\rangle + |1_0\rangle) =$$
$$= 2^{-\frac{n}{2}}(|0_{n-1}0_{n-2}...0_1 0_0\rangle + |0_{n-1}0_{n-2}...0_1 1_0\rangle + ... + |1_{n-1}1_{n-2}...1_1 1_0\rangle) \quad (13)$$

$|\psi\rangle$ belongs to the Hilbert space $H = H_{n-1} \otimes H_{n-2} \otimes ... \otimes H_0$. In our case where we have considered two anyons (the two qubits register):

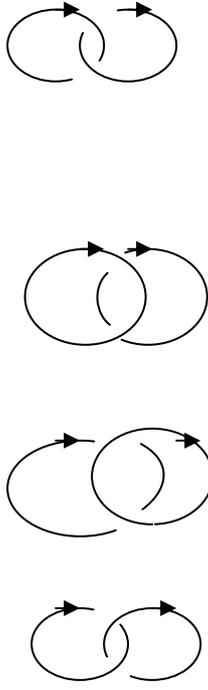

$$\begin{cases} |00\rangle = \begin{pmatrix}1\\0\end{pmatrix} \otimes \begin{pmatrix}1\\0\end{pmatrix} = \begin{pmatrix}1\\0\\0\\0\end{pmatrix} \\ |01\rangle = \begin{pmatrix}1\\0\end{pmatrix} \otimes \begin{pmatrix}0\\1\end{pmatrix} = \begin{pmatrix}0\\1\\0\\0\end{pmatrix} \\ |10\rangle = \begin{pmatrix}0\\1\end{pmatrix} \otimes \begin{pmatrix}1\\0\end{pmatrix} = \begin{pmatrix}0\\0\\1\\0\end{pmatrix} \\ |11\rangle = \begin{pmatrix}0\\1\end{pmatrix} \otimes \begin{pmatrix}0\\1\end{pmatrix} = \begin{pmatrix}0\\0\\0\\1\end{pmatrix} \end{cases}$$

Nearby, we have represented the corresponding quantum-2-link (for the + sign, we have considered the clockwise direction and for the – sign the opposite);.Topological feature shows us that $|00\rangle \neq |11\rangle$ and $|01\rangle = -|10\rangle$. On this base, with the Hamiltonian constant,

$$H = \frac{\pi \hbar}{2} \begin{pmatrix} 0 & 0 & 0 & 0 \\ 0 & 0 & 0 & 0 \\ 0 & 0 & 1 & -1 \\ 0 & 0 & -1 & 1 \end{pmatrix},$$

and with a well known unitary transformation called CNOT

$$U_{CNOT} = \begin{pmatrix} 1 & 0 & 0 & 0 \\ 0 & 1 & 0 & 0 \\ 0 & 0 & 0 & 1 \\ 0 & 0 & 1 & 0 \end{pmatrix},$$

we can map a quantum state (which is a product of two states) into an entangled state (hence, it can not be written as product of two states):

$$\frac{|0\rangle - |1\rangle}{\sqrt{2}} \otimes |0\rangle = 2^{-\frac{1}{2}} \begin{pmatrix} 1 \\ 0 \\ -1 \\ 0 \end{pmatrix} \xrightarrow{U_{CNOT}} \frac{|0\rangle - |3\rangle}{\sqrt{2}} = 2^{-\frac{1}{2}} \begin{pmatrix} 1 \\ 0 \\ 0 \\ -1 \end{pmatrix}$$

.The same correspondence can be represented as:

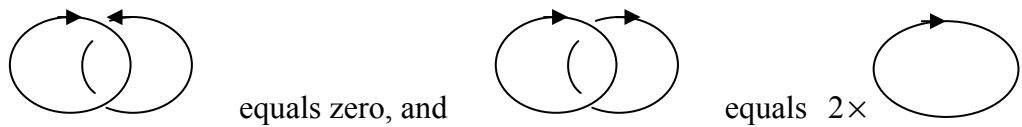

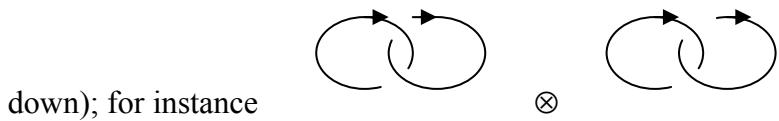

(14)

For these diagrams we are introducing some basic rules:

- the summation means to add and superpose the neighboring diagrams, giving them a well defined sign convention (for the + sign we have considered the clockwise direction and for the – sign, the opposite );for instance

   [diagram] equals zero, and [diagram] equals $2\times$ [diagram]

- the product means to concatenate on the vertical direction ( the right term, down); for instance [diagram] $\otimes$ [diagram]

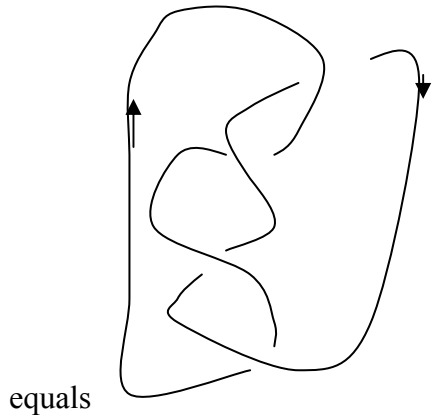

equals                   .

Using these rules on the correspondence (14), after some work, we get:

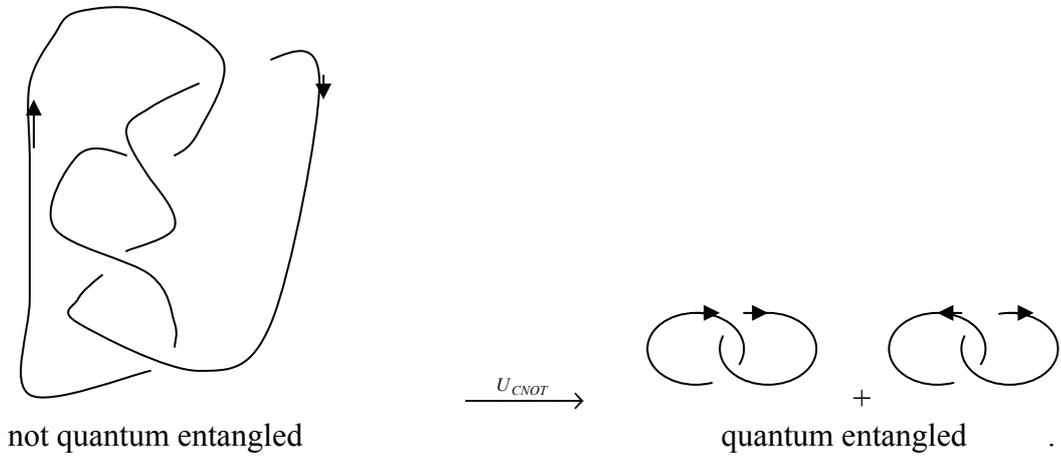

not quantum entangled                                quantum entangled             .

    The last diagram, that corresponds to the quantum entangled states, has an important feature: using the diagrammatic rules for summation, we'll always obtain the same thing. This may be written in this way:

    *A quantum state is entangled if it has an internal structure together with an operation (e.g. the summation) which leaves the structure intact.*

    If the number of the anyons is greater (e.g. three anyons as in Fig.6), the analyze is more difficult. In this cases a new feature arises, namely *the topological entropy*, which refers to the number of the quantum links corresponding to the same energy.

## *Conclusions*

    The quantum link theory is very promising; a lot of new features and evolutions have to be introduced in the future. In this paper, we have analyzed how the quantum-link appears through braiding. To do the same thing for the quantum knots seems to be more difficult since the anyons have to return in time on a specific segment of there road, otherwise we have to

introduce anti-anyons. Our subject, the quantum-links, will be further developed in the near future, analyzing some important features, such as *the $\rho$ distance and the tangle*. The importance of *the quantum link atlas* is obvious.